\documentclass[twocolumn,showpacs,preprintnumbers,a4,prl]{revtex4}
\usepackage{graphicx}%
\usepackage{dcolumn}
\usepackage{amsmath}
\usepackage{amssymb}
\usepackage{bm}

\makeatletter
\def\btt#1{\texttt{\@backslashchar#1}}%
\DeclareRobustCommand\bblash{\btt{\@backslashchar}}%
\makeatother

\begin{document}


\title{  Entanglements in Systems with Multiple Degrees of Freedom }
\author{Dong-Meng Chen$^{1,2}$, Wei-Hua Wang$^{1,2}$, and Liang-Jian Zou$^{1}$ }
\affiliation{ \it $^1$~  Key Laboratory of Materials Physics, Institute of Solid
                State Physics, Chinese Academy of Sciences, P. O. Box
                1129, Hefei 230031, China }
\affiliation{\it $^2$~ Graduate School of the Chinese Academy of Sciences}
\date{\today}


\begin{abstract}
In this letter we present the entanglement properties of the spin-orbital
coupling systems with multiple degrees of freedom.
After constructing the maximally entangled spin-orbital basis
of bipartite, we find that the quantum entanglement length in the
noninteracting itinerant Fermion system with spin and orbit
is considerably larger than that in the system with only spin.
In the SU(2)$\otimes$SU(2) spin-orbital interacting system,
the entanglement, expressed in terms of the spin-orbital correlation 
functions, clearly manifests the close relationship with the quantum phases
in strongly correlated systems; and the entanglement phase diagram
of the finite-size systems is in agreement with the magnetic and orbital
phase diagram of the infinite systems. 
The application of the present theory on nucleon systems is suggested. 

\pacs{03.65.Ud, 71.70.Ej, 73.43.Nq}

\end{abstract}
\maketitle

The total spin wavefunction of two entangled particles with spin-1/2
can not be expressed as the direct product of the wavefunctions
of the two individual particles, and the measurement to the spin of one
individual particle inevitably brings out the information of the spin states 
of the other particle. This property provides
wide potential application in quantum computer and quantum telepotation.
Entanglement properties in quantum many-particle systems have received great
interests in recent years since it is realized that, on the one hand,
the entanglement may characterize the quantum correlations
between particles \cite{Connor,Wang}, or even the quantum phase transitions 
in some simple models \cite{Osterloh, Vidal04,Gu03, Wu};
on the other hand, the realization of quantum computation in solid devices
also naturally raises the question, e.g. how these electrons entangle with
each other in the presence of strong correlations. Up to date, most of 
studies are concentrated on the spin entanglement of many-particle systems.
While in quantum many-particle systems, besides the spin
freedom degree, the particle may possess other degrees of freedom (DOF), such
as the orbit, the position and the momentum, etc. Especially in the strongly
correlated electronic systems, the orbital DOF is important for the relatively
localized 3d or 4f electrons \cite{Tokura}. Little was known about
the entanglement properties of the particles with multiple freedom degrees. 
In this letter we present the entangled properties of strongly correlated 
particles with multiple DOF. After constructing the maximally entangled 
states, the Bell basis, of two particles with spin and orbit, we first 
demonstrate that the quantum entanglement length in itinerant electron system 
with two DOF becomes significantly larger than that in the system with 
only spin. In the strongly interacting spin-orbital systems with 
SU(2)$\otimes$SU(2) symmetry, the entanglement evolves with the quantum 
phases, and appears abrupt changes near the critical points of the quantum
phase transitions. The entanglement phase diagram highly coincides with the 
magnetic and orbital phase diagram. Finally we discuss the possible 
application of the present theory on the nucleon systems with multiple DOF.
%

%
%
Considering the quarter-filling ($\nu$=1/4) quantum many-particle systems 
with spin ${\bf S}$=1/2 and twofold-degenerate orbital DOF, we introduce the 
pseudospin operator ${\bf \hat{\tau}}$ to describe the two orbital states:
${\bf \hat{\tau}}$=$\sum_{ab}C^{\dag}_{a}{\bf{\sigma}}_{ab}C_{b}$, here 
$\sigma$ is the Pauli matrix, and the orbital indices $a,b$ run over 1 and 2.
As well known, an entangled state is unfactorizable, and a
bipartite with maximally entangled state is completely
indivisibility. For a bipartite with two electrons at site A and
site B with spin DOF, the Bell basis consists of four states \(
|\psi_{1,2}^s \rangle_{AB}=\left(|\uparrow\downarrow \rangle
\pm |\downarrow\uparrow \rangle\right)/\sqrt{2}, \) and
$|\psi_{3,4}^s \rangle_{AB}=\left(|\uparrow\uparrow \rangle
\pm |\downarrow\downarrow \rangle\right)/\sqrt{2}$. These Bell
states with maximized entanglement are the foundation of quantum
information theory and quantum computation \cite{Nielsen}. 
In the presence of both spin and orbital DOF, we find that the possible 
Bell basis is separated into two independent groups.

(i) The first group is composed of the direct product of the entangled
spin and the entangled orbital parts, and each part is a Bell sub-basis,
i.e., maximally entangled. The 16 basis wavefunctions in the present 
situation can be readily constructed:
$|\psi^{s\tau}\rangle_{AB}=|\psi^{s}\rangle^{Bell}\otimes
|\psi^{\tau}\rangle^{Bell}$, where $|\psi^{s}\rangle^{Bell}$ and
$|\psi^{\tau} \rangle^{Bell}$ represent the spin and orbital Bell
states, respectively, e.g.
\begin{eqnarray}
|\psi^{s\tau}\rangle_{1,2}&=&\left(|\uparrow\uparrow \rangle \pm
|\downarrow\downarrow \rangle\right) \otimes \left( |11\rangle \pm
|22\rangle \right)
\end{eqnarray}
etc. Similar to the spin case, the 16 states in this group are the common 
eigenstates of the operator set
$(\sigma_{A}^x\sigma_{B}^x\tau_{A}^x\tau_{B}^x,
\sigma_{A}^z\sigma_{B}^z\tau_{A}^x \tau_{B}^x,
\sigma_{A}^x\sigma_{B}^x \tau_{A}^z\tau_{B}^z,
\sigma_{A}^z\sigma_{B}^z\tau_{A}^z\tau_{B}^z)$. 
One can use these operations to generate the 16 states in this group without 
difficulty.

(ii) The second group of the Bell basis is consisted of 16 spin-orbital states
which are the linear combination of independent spin-orbital parts, e.g.
\begin{eqnarray}
|\phi^{s\tau}\rangle_{1,2}&=&(|11\rangle^{\tau}\otimes|\psi_{1}^s\rangle
\pm|22\rangle^{\tau}\otimes|\psi_{2}^s\rangle)/\sqrt{2}
\end{eqnarray}
etc., here $|\psi_{i}^s\rangle$ is the $i$-$th$ spin Bell subbasis.
We notice that as the candidates for the maximally entangled basis, each 
group of these basis is orthonormalized and complete set.

In order to justify if these are the maximally entangled states, the
degree of entanglement of these states is quantified. 
%
%
For a many-state bipartite, the negativity is a good
operational measurement to quantify the entanglement since it is
monotonic under the local operation and the classical
communication, and it vanishes if two subsystems are not entangled
\cite{Vidal02}: $\aleph_{A,B}=\left( ||\rho^{T_{A}}||-1\right)/2$,
where $||\rho^{T_{A}}||$ is the trace norm of the partial
transpose of the density matrix of subsystem A versus to the B
and satisfies: $||\rho^{T_{A}}||=Tr\sqrt{\rho^{T_{A}}
(\rho^{T_{A}})^+}$. Interestingly, since spin and orbit are two
independent DOF, we could introduce the spin and the orbital
sub-entanglements to quantify the entanglement degree of each DOF
after trace out the other DOF: $\aleph^{s(\tau)}=(||\rho_{s(\tau)}
^{T_{A}}||-1)/2$, where $\rho_{s(\tau)}=Tr_{\tau(s)}(\rho)$.
%
%
We find both of the two groups of the basis have the maximal entanglement 
with the von Neumann entropy $E_{AB}=1$ and the negativity $\aleph_{AB}=1.5$. 
However the spin and the orbital sub-entanglements of these two subgroups
are different: in the first group, the spin or orbital maximal sub-entanglement 
of each state is $\aleph^{s(\tau)}_{AB}=0.5$;
whileas in the second group, the spin or the orbital sub-entanglement
of each state completely vanishes. Such significant difference in the
sub-entanglement reflects the distinct character of these two group basis:
the former is composed of the independent spin or orbital Bell states,
while the latter of spin-orbital indivisible bipartite.\\
%

%
%
\noindent{\it{\bf{A}}. Entanglement in Itinerant Fermion Systems}\\
Before investigating the entanglement in the spin-orbital interacting
system, we first explore the bipartite entanglement in
non-interacting itinerant electron system with spin and orbital
DOF. The spin entanglement of two electrons in noninteracting Fermi gas 
with single spin DOF was known \cite{Vedral03, Sangchul, Vedral05}.
In the system with only spin, the entanglement vanishes when
the spatial separation of two electrons is larger than a characteristic
length $r_{e}^0$, $r_{e}^0\approx 1.8/k_{F}^0$, where $k_{F}^0$ is the
zero-temperature Fermi momentum. 
In the spin-orbital itinerant electron system, we find that the
spin-orbital entanglement length of two electrons becomes considerably
large, and the  sub-entanglement of the bipartite for individual DOF is 
zero even in the most entangled spin-orbital states.

At zero temperature in the ground state, the electrons occupy all the 
levels below the Fermi surface
$|\psi_{0}\rangle=\prod_{s,\tau,|k|\le k_{F}} c_{{\bf k},s,\tau}^+|0\rangle$,
where the Fermi wavevector $k_{F}$ satisfies $k_{F}=(3\pi^2N/2V)^{1/3}$.
%
%
From the density matrix
$\hat{\rho_{N}}=|\psi_{0}\rangle\langle\psi_{0}|$, the two-particle
density matrix $\rho_{12}$ is obtained through tracing over
other particles
$\rho_{12}=Tr\{\rho_{N}\phi^+(2')\phi^+(1')\phi(1)\phi(2)\}$
\cite{yang}, where $1=({\bf r_{1}},\sigma_{1},\tau_{1})$, ${\bf
r_{1}}$ is the position vector, and the operator 
$\phi(1)=1/\sqrt{V}\sum_{{\bf k}}e^{-i{\bf
k\cdot r_{1}}} c_{{\bf k},s,\tau}$. Taking into account the spatial diagonal 
element, we obtain the reduced two-particle spin-orbital density matrix,
\begin{eqnarray}
\rho_{12}&=&[\delta_{\sigma_{1}\tau_{1};\sigma'_{1}\tau'_{1}}
\delta_{\sigma_{2}\tau_{2};\sigma'_{2}\tau'_{2}}-f(r)^2\delta_{\sigma_{1}
\tau_{1};\sigma'_{2}\tau'_{2}}\delta_{\sigma_{2}\tau_{2};\sigma'_{1}\tau'_{1}}]\nonumber \\
&&~~~~~~/[16-4f(r)^2]
\end{eqnarray}
where $f(r)=3(\sin{x}-x\cos{x})/x^3$ with $x=k_{F}r$.
For this mixed-state density matrix, the negativity is larger than zero as 
$f(r)> 1/2$,
\begin{equation}
\aleph_{12}=\frac{f(r)^2-\frac{1}{4}}{4-f(r)^2}.
\end{equation}
The negativity of two electrons as the function of the distance of
$k_{F}r$ at T=0 K for the spin-orbital and spin systems
is shown in Fig. 1. We find that in the present spin-orbital system
the quantum entanglement length $r_{e}$ is about $2.4/k_{F}$, considerably
larger than that in the spin system. Considering the shrinking of the Fermi
energy in the quarter-filling spin-orbital system, the entanglement
length is about 1.7 times larger than that in the spin system.
In the present itinerant electron system, the two-particle spin-orbital
density matrix (3) can be expressed as: 
$\rho=[4(1-f(r)^2)I/16+3f(r)^2\rho']/(4-f(r)^2)$,
where $\rho'$ is the density matrix composing of
all spin-orbital two-particle antisymmetric states and $I$
is the 16-order unit matrix. As a contrast, the
4-order density matrix in the spin system is
$\rho_{0}=[(2(1-f(r)^2)I_{0}/4+f(r)^2\rho'_{0}]/(2-f(r)^2)$,
where $\rho'_{0}$ is the density matrix of spin singlet
state and $I_{0}$ is the 4-component unit matrix.
Evidently, the unit matrix represents completely mixed disentangled 
state, so the more the weight of the unit matrix in $\rho$ is,
the more disentangled the state is. At sufficient large distance $k_{F}r$,
the weight of the unit matrix, or the disentangled component, in the 
spin-orbital density matrix $\rho$ is much smaller than that in the spin
density matrix $\rho_{0}$, leading to a significantly large quantum 
entanglement length in the spin-orbital system.

For two particles located at the same position, their spatial wavefunctions
are the same. Due to Pauli principle, the spin and orbital wavefunctions
must be antisymmetric, and it does not contributes to the unit matrix.
Thus the entanglement of the two electrons at is maximal at $r=0$, 
as shown in Fig.1.
Meanwhile, the spin-orbital wavefunction of the two electrons 
is composed of six antisymmetric spin-orbital substates, e.g.
$|\uparrow\uparrow\rangle\otimes |\psi^-_{\tau}\rangle$, etc., here
$|\psi^-_{\tau}\rangle $ is the orbital singlet state. These components
are not the Bell basis wavefunctions, so $\rho$, though is equal to $\rho'$, 
is still a mixed-state density matrix, and
the negativity is much smaller than the maximal value 1.5.
Comparing to the spin system, the entanglement length and the Fermi velocity
in the present system satisfy $r_{e}\approx 1.7 r_{e}^0$ and
$v_{F}=v_{F}^0/2^{\frac{1}{3}}$, respectively. Therefore, the coherent time
of the entangled electrons  $\tau_{c} \thicksim r_{e}/v_{F}$ is about 2.1 
times larger than that of the spin system. This property is also expected to 
valid for the localized electron systems.
Consequently, the present system with spin and orbit DOF is more favorable 
than the system with only spin 
in the realization of quantum computation and quantum telepotation .

\begin{figure}[htbp]\centering
\includegraphics[angle=270,width=0.8 \columnwidth]{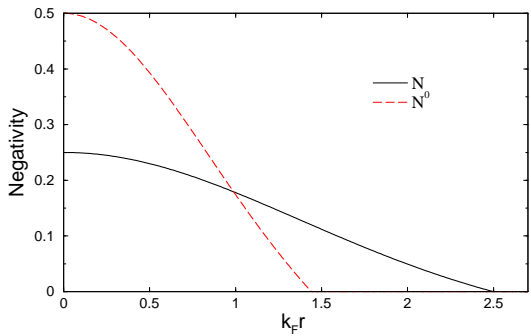}
\caption{Negativity as the function of spatial distance $k_{F}r$
of two electrons. $N$ and $N^0$ represent the negativity of two 
electrons in the spin-orbital system and in the spin system, respectively.}
\label{fig1}
\end{figure}
It is interesting to explore the variation of the spin sub-entanglement of
two electrons with the spatial distance. The reduced spin density matrix 
becomes,
\begin{equation}
\rho^s_{12}=\frac{1}{4-f(r)^2}[(4-2f(r)^2)\frac{I_{0}}{4}+f(r)^2|\psi_{s}^-
\rangle \langle\psi_{s}^-|]
\end{equation}
after tracing over the orbital DOF, where $|\psi_{s}^-\rangle$ is the spin
singlet state. We find that the spin-orbit density matrix of the two 
electrons at the same position, $\rho^s_{12}(r=0)$, is equal to that in the spin
system at the entangled distance, $\rho^0(r_{e})$, indicating that
even at the maximally entangled spin-orbital states,
the spin of two electrons in itinerant electron system is disentangled. 
The vanishing spin sub-entanglement in itinerant electron systems arises 
from the enlargement of Hilbert state space, which greatly reduces 
probability of forming spin singlet state.
The spin correlation functions, $\langle{\bf S_{i}S_{j}}\rangle$, also 
manifest the disentanglement of the spin DOF. We find at $r=0$, 
$\langle{\bf S_{i}S_{j}}\rangle=-1/4$, showing that 
the two electrons are classical AFM correlated. With the increase of
the separation between two electrons, the classical AFM correlation becomes 
smaller and smaller. Therefore the spin sub-entanglement in any spin-orbital 
state vanishes at any distance in itinerant electron systems.\\

\noindent{\it{\bf{B.}} Entanglement and Phase Diagram in Localized Electron 
Systems}\\
To explore the relationship between the variation of the entanglement
and quantum phase transitions in many-electron systems, we study the 
entanglement in the localized spin-orbital interacting systems in what follows.
Consider a one-dimensional system with strong spin-orbital correlation
\cite{Pati98, Yamashita}. Its Hamiltonian reads:
\begin{eqnarray}
H&=&\sum_{i}[{\bf S}_{i} \cdot {\bf S}_{i+1} {\bf \tau}_{i} \cdot
{\bf \tau}_{i+1}+J_{s}{\bf S}_{i} \cdot {\bf S}_{i+1}+J_{\tau}
{\bf \tau}_{i} \cdot {\bf \tau}_{i+1}\nonumber \\
&& +B_{s}S_{i}^z+B_{\tau}\tau_{i}^z]
\end{eqnarray}
where $J_{s}$ and $J_{\tau}$ are the spin and the orbital exchange
constants, respectively, and $B_{s}$ ($B_{\tau}$) is a small magnetic
(orbital) field. In the absence of the small external fields, both the spin 
and the orbital parts in Eq.(6) satisfy the SU(2) symmetry, the {\it so-called}
SU(2)$\otimes$SU(2) model. The periodic condition on Eq.(6) implies
that the entanglement between arbitrary two sites is an uniform function
of distance. The Hamiltonian commutes with the
z-components of the total spin and the total orbital operators,
$[H, S^z]=0$ and $[H, \tau^z]=0$, where $S^z=\sum_{i}S_{i}^z$
and $\tau^z=\sum_{i}\tau_{i}^z$. So each eigenstate of the Hamiltonian
is also the eigenstate of $S^z$ and $\tau^z$. One can use the index
($S^z$,$\tau^z$) to characterize the main four phases of this model
in different regions of the exchange constants $J_{s}$ and $J_{\tau}$:
Phase I with (S$^z$=N/2, $\tau^z$=N/2), Phase II with (N/2,0),
Phase III with (0,N/2) and Phase IV with (0,0), as shown in Fig.2.
In the present situation the spin-orbital entanglement can be explicitly
expressed in terms of the spin-orbital correlation functions.
Using the reduced density matrix of two nearest-neighbor particles,
one can analytically express the negativity $\aleph_{[i,i+1]}$ as the
nearest-neighbor spin-spin, orbital-orbital and spin-orbital correlation 
functions in each phase.
For instance in the phase IV, the negativity reads:
\begin{eqnarray}
\aleph_{[i,i+1]}&=&2|d|+|c-h|+|c+h|+|b-f|+|b+f| \nonumber \\
&&+\frac{1}{2}[|a-g-e+k|+|a-g+e-k|+|a \nonumber \\
&&+g-e-k|+|a+g+e+k|-1].
\end{eqnarray}
with
\begin{eqnarray}
a/b&=&\langle(\frac{1}{4}+S_{i}^zS_{i+1}^z)(\frac{1}{4} \pm 
\tau_{i}^z\tau_{i+1}^z)\rangle       \nonumber \\
c/d&=&\langle(\frac{1}{4}-S_{i}^zS_{i+1}^z)(\frac{1}{4} \pm
\tau_{i}^z\tau_{i+1}^z)\rangle        \nonumber \\
e/f&=&\langle(\frac{1}{4} \pm \tau_{i}^z\tau_{i+1}^z)(S_{i}^xS_{i+1}^x+
S_{i}^yS_{i+1}^y)\rangle \nonumber\\
g/h&=&\langle(\frac{1}{4} \pm S_{i}^zS_{i+1}^z)(\tau_{i}^x\tau_{i+1}^x+
\tau_{i}^y\tau_{i+1}^y)\rangle \nonumber\\
k&=&\langle(S_{i}^xS_{i+1}^x+S_{i}^yS_{i+1}^y)(\tau_{i}^x\tau_{i+1}^x+
\tau_{i}^y\tau_{i+1}^y)\rangle.
\end{eqnarray}
Obviously, in the fully polarized orbital phase III, the negativity has a 
simple form: $\aleph_{[i,i+1]}=-1/4-\langle{\bf S_{i}S_{i+1}}\rangle$.
Similarly we obtain the negativity in the completely polarized spin phase II,
$\aleph_{[i,i+1]}=-1/4-\langle{\bf\tau_{i}\tau_{i+1}}\rangle$.
The negativity vanishes in the fully polarized spin and orbital phase I.
Therefore, the direct relation between the quantum entanglement and the spin
and orbital correlations in strongly correlated electronic systems is thus
established, and it may provide many interesting information hidden in 
strongly correlated systems.

Thus one could clearly find that the variation of the entanglement closely
relates to the transition of quantum phases in the present system:
at the critical point of the quantum phase transition, the entanglement
exhibits a discontinuous change. The entanglement phase diagram of a 4-site
finite system is shown in Fig. 2. In the phase diagram Fig.2a, we adopt 
$\aleph_{[i,i+1]}$ and the sub-negativity $\aleph^{s(\tau)}_{[i,i+1]}$ 
to characterize these different quantum phases. 
\begin{figure}[htbp]\centering
\includegraphics[angle=270,width=1.0 \columnwidth]{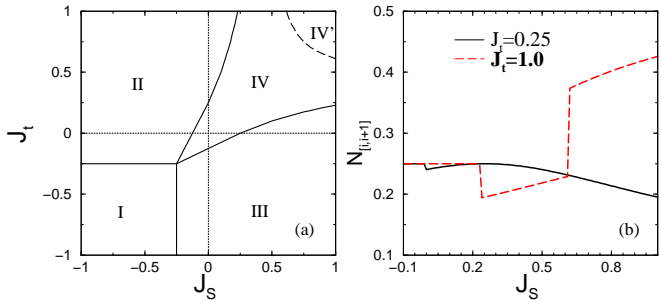}
\caption{Entanglement phase diagram of 4 particles in (a) and
dependence of negativity on magnetic exchange constant $J_{s}$
at $J_{\tau}=0.25$ and $J_{\tau}=1.0$ in (b). Phases I, II, III
and IV represent the FM and FO state with ($S^z$=2, $\tau^z$=2),
the FM and AFO state with (2,0), the AFM and FO state
with (0, 2), and the AFM and AFO state with (0, 0), respectively.
Phase IV differs from IV$'$ in the spin-orbital correlation function,
see the address in the text.}
\label{fig2}
\end{figure}
Very interesting, we find that the finite-site entanglement phase diagram in 
Fig. 2(a) is
almost identical to the magnetic and orbital phase diagram in the 
one-dimensional infinite lattice \cite{Pati98, Yamashita}, indicating that 
the quantum information of infinite systems manifests through the 
entanglement of proper finite clusters.
The phase diagram in Fig.2a can also be characterized by the spin-orbital 
index ($S^z$, $\tau^z$).
Among these phases in Fig.2a, the states in Phase I are disentangled,
corresponding to fully polarized spin and orbital states with ($S^z$=2, 
$\tau^z$=2), i.e. the ferromagnetic (FM) and ferro-orbital (FO) state.
In Phase II, the negativity satisfies
$\aleph_{[i,i+1]}$=$\aleph^{\tau}_{[i,i+1]}$ and $\aleph^{s}_{[i,i+1]}$=0,
indicating the states in Phase II are FM and antiferro-orbital (AFO) 
with (2,0). Similarly, Phase
III corresponds to the AFM and FO states with (0,2). Both Phases IV and 
IV$'$ correspond to AFM and AFO with (0,0), while these two phases are
distinguished by the spin-orbit correlation functions. In Phase IV,
$\langle{\bf S}_{i} \cdot {\bf S}_{i+1} {\bf \tau}_{i}\cdot{\bf \tau}_{i+1} 
\rangle$$<$0, and in Phase IV$'$, $\langle{\bf S}_{i}
\cdot{\bf S}_{i+1} {\bf \tau}_{i} \cdot {\bf \tau}_{i+1}\rangle$$>$0.
This difference arises from the fact that when either $J_{s}$ or $J_{\tau}$
becomes large while the other one is small,
$\langle {\bf S}_{i} \cdot {\bf S}_{i+1} \rangle<0$ and
$\langle {\bf \tau}_{i}\cdot{\bf \tau}_{i+1} \rangle<0$.
Due to the quantum fluctuation, $\langle{\bf S}_{i} \cdot {\bf S}_{i+1} {\bf 
\tau}_{i}\cdot{\bf \tau}_{i+1} \rangle<0$, in Phase IV. When both $J_{s}$ 
and $J_{\tau}$ become large enough, the strong AFM and AFO correlations 
polarize the spins and orbits, and lead the spin-orbital correlation function
$\langle{\bf S}_{i} \cdot {\bf S}_{i+1}{\bf \tau}_{i}\cdot{\bf \tau}_{i+1} 
\rangle$ to transition from negative to positive, as we see the phase IV$'$ 
in Fig.2a. 
The discontinuous changes of entanglement of two electrons near the phase 
boundary are in accordance with the critical points of the quantum phase
transitions, as clearly shown in Fig 2(b). At J$_{s}$=J$_{\tau}$=1/4 and 
B$_{s}$=B$_{\tau}$=0, the system exhibits SU(4) symmetry \cite{Liyouquan},
then the ground state is highly degenerate and the negativity comes to the 
maximum. These results clearly demonstrate that the entanglement is closely 
related with the quantum phase transitions in strongly correlated systems.
 It is worthy of pointing out that we introduce the small magnetic and
orbital fields to lift the degeneracy of Phases II and III. In the
absence of these small fields the entanglement is uncertain, as discussed
by Qian, et al \cite{Qian}.

It is interesting that the present theory is also applicable for the nuclear systems,
in which the nucleons usually possess more than one DOF, such as spin, isospin, etc.
According to our results for the itinerant Fermion systems, the
entanglement length of the nucleons is significantly larger than one expects.
We anticipate more interesting entanglement properties in the systems with 
multiple degrees of freedom will be uncovered in the further studies.

%

   Authors thanks for the useful discussion with Z. S. Ma.
Supports from the NSF of China and the BaiRen Project from
the Chinese Academy of Sciences (CAS) are appreciated. 
Numerical work was performed in CCS, HFCAS.


\end{document}